\documentclass[preprint,showpacs,amsmath,amssymb,aps]{revtex4}

\usepackage{graphicx}
\usepackage{dcolumn}
\usepackage{bm}

\begin{document}

\title{Critical Current Calculations For Long $0$-$\pi$ Josephson
Junctions}
%CHANGE: I suggest a change to the title to reflect the fact that this is more of a
%model calculation than an accurate reproduction of the experiments.

\author{I.\ Tornes and D.\ Stroud}

\affiliation{Department of Physics, Ohio State University, Columbus, OH 43210}

\date{\today}

\begin{abstract}

A zigzag boundary between a $d_{x^2-y^2}$ and an $s$-wave
superconductor is believed to behave like a long Josephson junction
with alternating sections of $0$ and $\pi$ symmetry.  We calculate the
field-dependent critical current of such a junction, using a simple
model.  The calculation involves discretizing the partial differential
equation for the phase difference across a long $0$-$\pi$ junction.  In this form, the equations
describe a hybrid ladder of inductively coupled small $0$ and $\pi$
resistively and capacitively shunted Josephson junctions (RCSJ's).  
The calculated critical critical current density
$J_c(H_a)$ is maximum at non-zero applied magnetic field $H_a$, and
depends strongly on the ratio of Josephson penetration depth
$\lambda_J$ to facet length $L_f$.  If $\lambda_J/L_f \gg 1$ and the
number of facets is large, there is a broad range of $H_a$ where
$J_c(H_a)$ is less than $2\%$ of the maximum critical current density
of a long $0$ junction.  All of these features are in qualitative
agreement with recent experiments.  In the limit $\lambda_J/L_f
\rightarrow \infty$, our model reduces to a previously-obtained
analytical superposition result for $J_c(H_a)$.  In the same limit, we
also obtain an analytical expression for the effective field-dependent
quality factor $Q_J(H_a)$, finding that $Q_J(H_a) \propto
\sqrt{J_c(H_a)}$.  We suggest that measuring the field-dependence of $Q_J(H_a)$
would provide further evidence that this RCSJ model applies to a long $0$-$\pi$ junction
between a d-wave and an s-wave superconductor.

\end{abstract}


\maketitle

\section{Introduction}

Most hole-doped cuprate-based high-temperature superconductors are
believed to be characterized by a d$_{x^2-y^2}$ order parameter.  Such
an order parameter leads to many measurable consequences, but perhaps
the most dramatic are due to so-called $\pi$ junctions between a
$d_{x^2-y^2}$ superconductor and either a conventional, s-wave
superconductor or another $d_{x^2-y^2}$ superconductor.  These $\pi$
junctions can be detected by phase-sensitive symmetry
measurements\cite{rmp}.  In a $\pi$ junction, the Josephson current
$I$ and the gauge-invariant phase difference $\gamma$ across the
junction are related by $I = I_c\sin(\gamma + \pi)$\cite{tinkham},
where $I_c > 0$ is the junction critical current.  By contrast, a
conventional $0$ junction satisfies $I = I_c\sin\gamma$.  When a
Josephson loop is formed by a suitable arrangement of d-wave and
s-wave superconductors, the current-voltage (IV) characteristics
differ dramatically from those found when only s-wave superconductors
are involved.  Such phase-sensitive experiments have provided some of
the most persuasive evidence that the high-T$_c$ superconductors
indeed have a d$_{x^2-y^2}$ superconductor order parameter.

Recently, the d-wave nature of the order parameter was further
confirmed by measurement of the IV characteristics of long ``zigzag''
junctions connecting two superconductors: a cuprate superconductor
 with an order parameter of $d_{x^2-y^2}$ symmetry, and Nb, an s-wave
superconductor\cite{darminto,smilde}.   In the experiments of Ref.\
\onlinecite{darminto}, the cuprate superconductor was YBa$_2$Cu$_3$O$_{7-x}$ (YBCO),
while in Ref.\ \onlinecite{smilde}, Nd$_{2-x}$Ce$_x$CuO$_{4-y}$ (NCCO) was used.  
The geometry of the zigzag junction is shown in Fig.\ 1 of Ref.\ \onlinecite{darminto}, which also
shows the expected orientation of the order parameter lobes on each
side of the junction.  In this configuration, the zigzag junction is
expected to consist of alternating sections of $0$ and $\pi$
junctions, as shown in that figure.  In the experiments, the IV
characteristics of this zigzag junction were measured in the presence
of an external magnetic field applied parallel to the facets of the
superconductor,{\it i. e.} perpendicular to the plane of the figure. The
response of the junctions was, indeed, found to be consistent with the
expected $d_{x^2-y^2}$ symmetry of YBCO and NCCO.

In this paper, we use a simple numerical model to calculate the
critical current density $J_c(H_a)$ of a long $0$-$\pi$ Josephson
junction as a function of the applied magnetic field $H_a$. One of our
goals is to obtain a general picture of how $J_c(H_a)$ behaves in this
model as a function of the variables $\lambda_J/L_f$ and $N_f$, where
$L_f$ is the length of one facet of the junction $\lambda_J$ is the
Josephson penetration depth, and $N_f$ is the number of facets.  A
second goal is to see how well our simple model agrees with the
experimental results\cite{darminto,smilde} for $0$-$\pi$ zigzag
junctions.  A final goal is to connect our approach to other
treatments of long $0$-$\pi$ junctions.  We find that, in the limit
$\lambda_J/L_f \gg 1$, the model reduces to a generalized
superposition approximation previously used to treat such
junctions\cite{smilde,mints}.  However, even in that limit, we obtain
new information, namely, an analytical expression for the
field-dependent quality factor $Q_J(H_a)$ of the long junction.

A characteristic feature of a one-dimensional array of alternating $0$
and $\pi$ junctions is the occurrence of half-integer flux quanta, in
the region where the $0$ and $\pi$ junctions join. Many features of
these half-integer flux quanta have been studied analytically and
numerically by Goldobin {\it et al.}~\cite{goldobin,goldobin1}, in
zero applied magnetic field.  Our approach is basically a discretized
version of this model, but extended to a finite magnetic field.  In
particular, our model does contain the half-quanta which are
responsible for some of the observed experimental features.

Qualitatively, our calculated IV characteristics agree fairly well
with experiment.  For example, the measured field-dependent critical
current $I_c(H_a)$ of the zigzag junction found to be symmetric in
$H_a$, with a principal maximum at nonzero $H_a$. This behavior is in
contrast to a long $0$ (or long $\pi$) junction, which will have a
maximum $I_c(H_a)$ at $H_a = 0$. Our model reproduces these features.
We also obtain the result that, when $\lambda_J/L_f \gg 1$ and the
long junction has many facets, there is a broad range of $H_a$ where
$J_c(H_a)$ is only $\sim 1$-$2$\% of the maximum critical current
density.  In the experiments of \cite{darminto,smilde}, it is found that
$J_c(H_a)$ is very small over a wide range of $H_a$.

The remainder of this paper is arranged as follows.  In Section II, we
describe our simple model.  The numerical results obtained from this
model are given in Section III.  In Section IV, we show that the model
reduces to a generalized superposition approximation in the limit
$\lambda_J/L_f \rightarrow \infty$, and we obtain the effective
quality factor $Q_J(H_a)$ in the same limit.  Finally, we give a
concluding discussion in Section V.

\section{Formalism}

We consider a long Josephson junction consisting of alternating $0$
and $\pi$ sections, or facets, as in the zigzag junctions studied
experimentally. Let the length of one such facet be $L_f$, and let
there be $N_f$ such facets, so that the total junction length is $L =
N_fL_f$.  We use coordinates such that the alternating series of
facets runs along the $x$ axis, and the plates of the junction are
normal to the $y$ axis.  In addition we assume that there is a uniform
magnetic field ${\bf H}_a = H_a{\bf \hat{z}}$ applied parallel to the
junction plates.

%CORRECTION: I corrected the above formula, which was wrong in the original.

The gauge-invariant phase $\gamma(x, t)$ across the junction satisfies
the following partial differential equation:
\begin{equation}
\gamma_{tt} = - \sin\gamma + \lambda_J^2\gamma_{xx} + \frac{J}{J_{c0}}
+ \frac{1}{Q_J}\gamma_t - \pi \lambda_J^2n_{xx}. \label{eq:singord2}
\end{equation}
Eq.\ (\ref{eq:singord2}) is equivalent to eq.\ (1) of Ref.\
\onlinecite{goldobin1} in the case when $H_a$ is uniform.  In this equation,
the subscripts $x$ and $t$ represent derivatives with respect to
position $x$ and a dimensionless time $\omega_pt$, $x$ being the
coordinate along the junction.  $\lambda_J$ is the Josephson
penetration depth and $\omega_p$ is the Josephson plasma frequency,
$J$ is the driving current per unit length across the junction
(assumed uniform), and $J_{c0}$ is the critical current per unit
length of a $0$ junction at zero applied field ($H_a = 0$).  In terms
of the junction parameters, $\omega_p^2 = 2e J_{c0}/(\hbar C)$ where
$C$ is the junction capacitance per unit length. $Q_J = \omega_p C/G$
is the junction quality factor, and $G \equiv 1/R$ is the normal shunt
conductance of the junction per unit length. For a junction connecting
two non-identical materials $\lambda_J^{-2} = 8\pi e (\lambda_1 +
\lambda_2 + d)\tilde{J}_c/(\hbar c^2)$\cite{tinkham}, where
$\tilde{J}_c$ is the critical current per unit {\it area} and
$\lambda_1$ and $\lambda_2$ are the penetration depths of the
materials on either side of the junction.  Finally, $n(x)$ indicates
whether the point $x$ lies within a $0$ or a $\pi$ facet: $n(x) = 0$
in a $0$ facet and $n(x) = 1$ in a $\pi$ facet.
%IVAN6: I believe that, in this form, the actual (as opposed to applied) magnetic
%field need not be uniform.  Do you agree?
%STROUD6:  I'm honestly not sure I understand what you mean by the actuaal (as opposed
% to applied) field.  

In this formulation, the applied magnetic field does not appear
explicitly in the differential equation; instead, it appears as a
boundary condition: $\gamma(L, t) - \gamma(0, t) = 2\pi H_a
d_{\mathrm{eff}}L/\Phi_0 + [n(L) - n(0)]\pi$, where $\Phi_0 = hc/(2e)$
is the flux quantum.

The gauge-invariant phase can also be expressed as
\begin{equation}
\phi(x, t) = \gamma(x, t) - \frac{2\pi}{\Phi_0}\int_i^j{\bf A}_a\cdot {\bf dl} -
\pi n(x),
\end{equation}
where ${\bf A}_a$ is the vector potential corresponding to the applied
field ${\bf H}_a$, and the integral runs between the plates of the
junction in the $y$ direction.  It is convenient to work in a gauge
such that ${\bf A}_a = H_ax\hat{y}$, so that $(2\pi/\Phi_0)\int{\bf
A}\cdot {\bf dl} = 2\pi H_a d_{\mathrm{eff}}
x/\Phi_0$. $d_{\mathrm{eff}}$ is the effective thickness of the junction,
which we may express in terms of
$\lambda_1$, $\lambda_2$, and the actual Josephson barrier thickness $d$ as
$d_{\mathrm{\mathrm{eff}}} = d + \lambda_1 + \lambda_2$.
%STROUD6:  I don't understand this sentence
%IVAN6.  I replaced d by d_{\mathrm{eff}} and introduced \lambda_1 and \lambda_2.
%What do you think?\
%STROUD6:  I think that is fine, but I did not consider them when I did my calculations
%So the d that I used is some combination of \lambda_1 and \lambda_2
%IVAN7: I corrected the above sentence so it now makes sense, I think. 
 
The equation of motion for the new variable $\phi(x, t)$ across
the junction may now be written as
\begin{equation}
\phi_{tt} = -\sin\left[\phi + \frac{2\pi}{\Phi_0}H_a
d_{\mathrm{eff}} x + \pi n(x)\right] + \lambda_J^2\phi_{xx} +
\frac{J}{J_{c0}} - \frac{1}{Q_J}\dot{\phi}, \label{eq:singord1}
\end{equation}
with the boundary conditions $\phi_x(0,t) = \phi_x(L,t) = 0$.
For numerical purposes, eq.\ (\ref{eq:singord1}) is generally more
convenient than eq.\ (\ref{eq:singord2}), because, in the latter,
$n(x)$ is only piecewise continuous, and hence $n_{xx}$ is the sum of
derivatives of Dirac delta functions.  Note also that $H_a$ appears
explicitly in eq.\ (\ref{eq:singord1}), whereas in eq.\
(\ref{eq:singord2}) it appears only as a boundary condition.
 
In order to treat eq.\ (\ref{eq:singord1}) numerically, it is
convenient to discretize it in space, by breaking the long junction
into $N$ small segments, or ``mini-junctions,'' each of length
$\Delta$ ($L = N\Delta$). We also assume that each $0$ or $\pi$ facet
is divided into an integer number $N_p$ of such mini-junctions, so
that $L_f = N_p\Delta$ and thus $N_pN_f = N$.  Denoting the phase
difference across the i$^{th}$ such junction by $\phi_i$, we obtain a
collection of $N$ coupled ordinary differential equations.  Except for
$i = 1$ and $i = N$, these may be written
\begin{equation}
\ddot{\phi}_i = -\sin(\phi_i+2\,\pi i\,f/N_p + n_i\pi) +
\frac{\lambda_J^2}{\Delta^2}(\phi_{i+1}-2\phi_i+\phi_{i-1}) +
\frac{J}{J_{c0}} - \frac{1}{Q_J}\dot{\phi}_i. \label{eq:phiddot}
\end{equation}
\noindent
%Here the last term corresponds to the junction quality
%factor $Q_J = \omega_p {\cal C}/{\cal G}$ , and has been added by
%hand to the equation in order to account for shunt conductance of
%the junction.  ${\cal G}$ is the shunt conductance per unit length
%of the long junction. Expressing the shunt conductance of an
%individual mini-junction as ${\cal G}\Delta = 1/R$, and the
%mini-junction capacitance as ${\cal C}\Delta = C$, we see that the
%value of $Q_J$ is independent of the choice of $\Delta$, and may
%also be written $Q_J = \omega_pRC$.  We may also express $\omega_p$
%in terms of parameters of one mini-junction as $\omega_p =
%[2eI_c/(\hbar C)]^{1/2}$.
Here we have introduced the frustration $f = H_a L_f
d_{\mathrm{eff}}/\Phi_0$, defined as the flux per facet in units of
$\Phi_0$.
%CORRECTION: I have replaced d by d_{\mathrm{eff}} here and elsewhere.
For $i = 1$ and $i = N$, the second term on the right
hand side of eq.\ (\ref{eq:phiddot}) must be modified, and the equation of motion becomes
\begin{equation}
\ddot{\phi}_i =  -\sin(\phi_i+2\,\pi i\,f/N_p+n_i\pi) +
\frac{\lambda_J^2}{\Delta^2}(\phi_{i+1}-\phi_i) +
\frac{J}{J_{c0}}\phi_i - \frac{1}{Q_J}\dot{\phi}_i, \label{eq:phiddot1}
\end{equation}
\noindent and
\begin{equation}
\ddot{\phi}_i = -\sin(\phi_i+2\,\pi i\,f/N_p+n_i\pi) +
 \frac{\lambda_J^2}{\Delta^2}(\phi_{i-1}-\phi_i) + \frac{J}{J_{c0}}\phi_i - \frac{1}{Q_J}\dot{\phi}_i
\label{eq:phiddotn}
\end{equation}
\noindent for $i = 1$ and $i = N$ respectively.

Eqs.\ (\ref{eq:phiddot}) - (\ref{eq:phiddotn}) describe a {\em hybrid
Josephson ladder}, consisting of N mini-junctions inductively coupled
together in parallel, and driven by an applied uniform dc current.
The geometry is shown schematically in Fig.\ 1.  The driving current
is in the $y$ direction and the field is in the $z$ direction.  We use
the phrase ``hybrid ladder'' to distinguish this system from the more
conventional Josephson ladders, which have junctions on both the rungs
and the edges\cite{flach}.  The limit of physical interest, however,
is the continuum limit, i.\ e., $N_p \rightarrow \infty$.

\section{Numerical Results}

Before describing our results, we make some qualitative remarks about
the relation between our model and the zigzag junction used in
experiments\cite{darminto,smilde}.  Although our $0$-$\pi$ junction is
straight, it does have one crucial feature in common with the zigzag
junction, namely, intrinsic frustration, even at zero applied magnetic
field.  We define the frustration $\tilde{f}_i$ of the plaquette lying
between the i$^{th}$ and $(i+1)^{th}$ mini-junction as
\begin{equation}
\tilde{f}_i = \frac{f}{N_p}
\end{equation}
if both junctions are either $0$ junctions or $\pi$ junctions, and
\begin{equation}
\tilde{f}_i = \frac{f}{N_p} + \frac{1}{2}
\end{equation}
if one is a $0$ junction and the other is a $\pi$ junction.  If
$\tilde{f}_i$ is an integer, the energies of all the bonds making up
the plaquette can be simultaneously minimized, and the plaquette is
unfrustrated.  If $\tilde{f}_i$ is non-integer, the bond energies
cannot all be simultaneously minimized, and the actual bond
configuration is a compromise among these; in this case, the plaquette
is frustrated.

%IVAN5: I redefined frustration so that it means frustration per
%facet, not per plaquette.  Also, in the next two paragraphs, I try
%to give a qualitative picture of how frustration affects the
%critical current.

The effects of frustration are easily understood if $N_p = 1$.  Then
the long $0$-$\pi$ junction is represented by a hybrid ladder having
alternating sections of $0$ and $\pi$ junctions.  Thus, according to
the above definition, the ladder is frustrated ($\tilde{f}_i \neq 0$)
at all fields, including $H_a = 0$, with the exception of $f = 1/2$,
i. e., one-half quantum of flux per facet.  At this field, the ladder
is unfrustrated, and we expect that the critical current per facet
will be maximum.  Moreover, that maximum (per facet) should equal the
critical current of a single $0$ facet at $H_a = 0$, namely
$L_fJ_{c0}$.  At all other fields, the critical current will be
smaller than that maximum value.

If $N_p > 1$, the ladder has some residual frustration even at $f =
1/2$.  For example, at $N_p = 2$ and $f = 1/2$, each plaquette has
$\tilde{f}_i = 3/4$.  Thus, although we expect the critical current
still to be maximum at $f = 1/2$, that maximum will be smaller than
$L_fJ_{c0}$.
%The physics of this residual frustration is that, when
%$N_p > 1$, the phase $\phi(x)$ can vary along the facet.
%IVAN5: I removed the above sentence because there is frustration even
%if the phase does not vary along the facet.
Since the physical system is really in the continuum limit ($N_p
\rightarrow \infty$), we generally expect a critical current which is
smaller than $L_fJ_{c0}$ even at the optimum value of $f = 1/2$.  This
expectation is borne out by our calculations below.

\begin{figure}
\begin{center}
\includegraphics[width=0.75\textwidth]{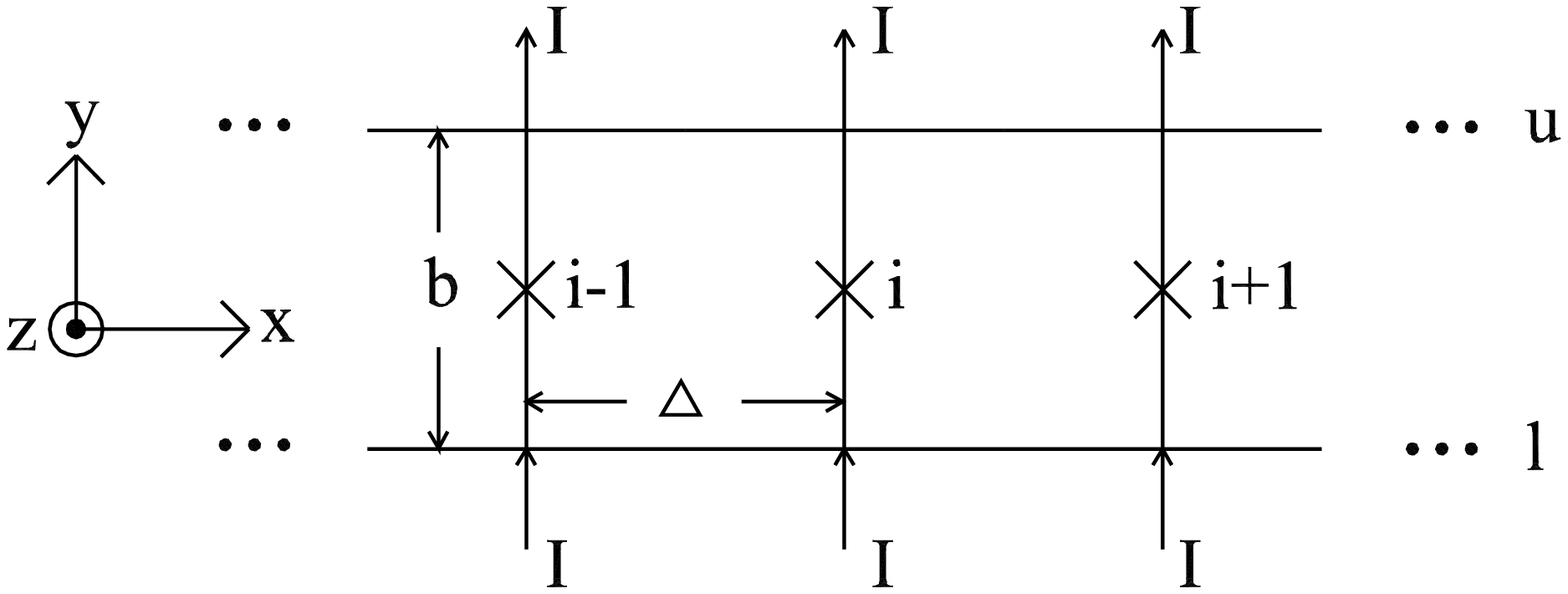}
\caption{ \small Schematic view of Josephson ladder used in present
calculations. The ladder has open boundary conditions; $u$ and $\ell$
represent the upper and lower edges of the ladder. A uniform current
density $J$ per unit length is applied in the $\hat{y}$ direction.
This corresponds to a current $I = J\Delta$ applied across each rung
in the ladder.  An external magnetic field ${\bf H}_a$ is applied in
the $\hat{z}$ direction. }\label{ladder_fig}
\end{center}
\end{figure}

Although the $0$-$\pi$ ladder configuration of Fig.\ \ref{ladder_fig}
does have the frustration characteristic of a zigzag junction, it
still differs geometrically from the one shown in Fig.\ 1 of Ref.\
\onlinecite{darminto}, even in the continuum limit, because the zigzag
junction is not straight.  The bends in the zigzag junction will
introduce some pinning which is absent from our model, and therefore
will probably lead to slightly different IV characteristics than those
produced by the model of Fig.\ \ref{ladder_fig}, possibly more closely
resembling experiment.
%IVAN6: note the changes in the above paragraph.  I also thought we
%could remove the Figure with the extra plaquettes, since we do not
%carry out any calculations in this geometry.  What do you think?
%STROUD6:  I'm fine with that.  It probably only confuses people having it 
%in there.  So, taking it out is good. 

%CHANGE: Add paragraph below in response to second report of first referee.

We turn now to our calculations. We have solved eqs.\
(\ref{eq:phiddot}) - (\ref{eq:phiddotn}) for $0$-$\pi$ junctions made
of various numbers of facets $N_f$, numbers $N_p$ of mini-junctions
per facet, facet lengths $L_f$, Josephson penetration depths
$\lambda_J$, and applied fields $f$. In all cases, we obtain the
solutions to the coupled differential equations using a standard
fourth order Runge-Kutta technique.  Our goal is to obtain the
critical current of a long $0$-$\pi$ junction as a function of these
parameters, as well as of the Josephson penetration depth $\lambda_J$,
In principle, we are interested in the continuum limit, corresponding
to $N_p \rightarrow \infty$.  Any effects arising from finite $N_p$
will be absent from a real $0$-$\pi$ junction, and are therefore
artifacts of the finite discretization length $\Delta$.  In general,
we expect that $\phi(x)$ will vary only on a length scale of
$\lambda_J$.  Therefore, if $L_f \ll \lambda_J$, we should not need a
large $N_p$ to approach this continuum limit.  Our numerical results
support this expectation, as we discuss below.
%In most of our calculations, we
%attempt to choose parameters resembling the experimental
%ones\cite{darminto,smilde}, though we also consider more general
%cases.
%IVAN5: the above sentence is repeated below, I believe.

We first consider a single $0$ junction in an external field $f$. If
$\lambda_J > L$, then the critical current per unit length, $J_c(f)$,
should exhibit a Fraunhofer-like pattern\cite{tinkham}.  Fig.\ 2 of
Ref.\ \onlinecite{darminto}, which shows their measured $I_c(H_a)$ for a
single long $0$ junction, clearly exhibits this Fraunhofer pattern.

To calculate this pattern in our model, we apply a specified field $f$
to the junction, then ramp up the external driving current density $J$
until a non-zero voltage is recorded, giving $J_c(f)$.  $f$ is then
increased and the process is repeated.  Since the value of $Q_J$ is
not explicitly given in the experimental papers\cite{darminto,smilde},
we arbitrarily choose $Q_J = 6$ in all our calculations. This value
will lead to hysteretic IV characteristics for at least some $f$, as
in the experiments.

\begin{figure}
\begin{center}
\includegraphics[width=0.75\textwidth]{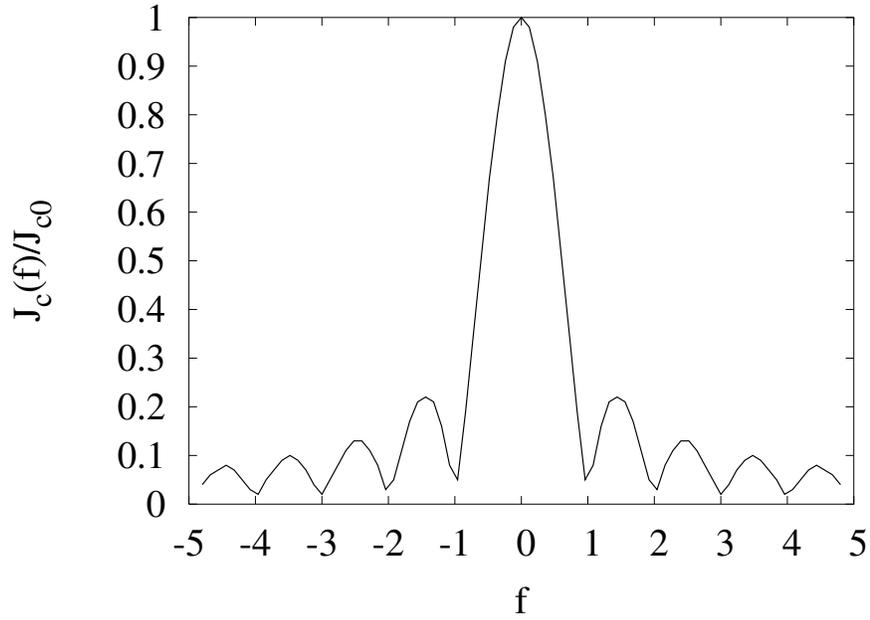}
\caption{\small Calculated critical current per unit length $J_c(f)$
for a $0$ junction of length $L_f$, plotted as function of the flux
$f$ through the facet.  $f$ is given in units of a flux quantum
$\Phi_0 = hc/(2e)$.  The facet is represented as a hybrid ladder of
120 small $0$ junctions, as described in the text.  The Josephson
penetration depth $\lambda_J = 1.3L_f$ and the quality factor $Q_J =
6.0$.  $J_{c0}$ is the critical current per unit length of a $0$
junction at zero applied field.}\label{fraun_ivan}
\end{center}
\end{figure}

In Fig.\ \ref{fraun_ivan}, we show our calculated $J_c(f)/J_{c0}$ for
a hybrid Josephson ladder of $120$ rungs with all $0$ junctions, open
boundary conditions, a penetration depth $\lambda_J = 1.3 L$, and $Q_J
= 6$.  Fig.\ ~\ref{fraun_ivan} shows that our ladder configuration
behaves similarly to the long $0$ junction studied experimentally
(Fig.\ 2 of Ref.\ \cite{darminto}), and hence, our discretization of
the long junction into 120 mini-junctions is on a sufficiently fine
scale to produce the Fraunhofer pattern seen experimentally.
Comparing Fig.\ \ref{fraun_ivan} to the experimental Figure, we see
that $f \sim 1.44$ is equivalent to $H_a \sim 1.1 \mu$T.  Since all
our IV characteristics are obtained on the increasing current branch,
we have not checked for the expected hysteresis seen in the
experiment.
%IVAN5: I redefined f so that it is the field per facet, in units
%of \Phi_0

According to conventional theory\cite{tinkham}, the Fraunhofer pattern
in a long Josephson junction results from simple superposition:
$J_c(H_a) = J_{c0}|\sin(\pi \Phi/\Phi_0)/(\pi \Phi/\Phi_0)|$, where
$\Phi$ is the flux through the junction area $L(d + 2\lambda)$.  This
result is obtained\cite{tinkham} by assuming that the phase $\phi(x)$
[eq.\ (2)] is {\em independent} of x, or equivalently, that $\phi_i$
is independent of $i$ [eqs.\ (4) - (6)]. Although, for this particular
$\lambda_J/L_f$, we obtain numerical results very close to this
asymptotic form, our approach is more general than simple
superposition.  In particular, we do {\em not} assume that the
$\phi_i$'s are all the same.  Instead, the $\phi_i$'s, at any applied
current density $J$, are always the values which satisfy the coupled
differential equations. Thus, in principle, our approach gives a
different result from superposition, even for this simple case.  This
point is discussed further below.
%IVAN5: the above paragraph is new.  What do you think?  Do we need
%to say anything more?

%Fig.\ 3 of Ref.\ \cite{darminto} shows the measured $I_c(H_a)$ for a
%zigzag junction consisting of eight $25 \mu$m facets of alternating
%$0$ and $\pi$ long junctions.  The measured $I_c(H_a)$ is symmetric
%about $H_a = 0$.

%We have calculated $I_c(f)/I_c(0)$ for various ladder configurations
%and obtained rather similar results.

\begin{figure}
\begin{center}
\includegraphics[width=0.75\textwidth]{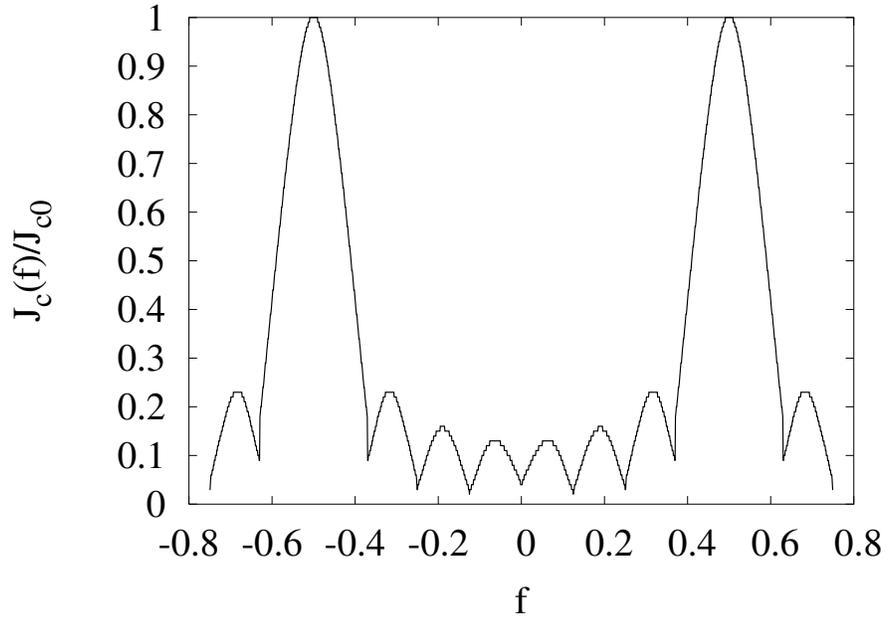}
\caption{\small Critical current per unit length $J_c(f)$, in units of
$J_{c0}$, plotted as a function of the magnetic flux per facet $f$,
for an eight-facet junction of alternating $0$ and $\pi$ facets, open
boundary conditions, $\lambda_J=2.6L_f$ and $Q_J = 6.0$.  The junction
is modeled as a hybrid Josephson ladder with one mini-junction per
facet, as described in the text.}\label{eight_1mini}
\end{center}
\end{figure}

\begin{figure}
\begin{center}
\includegraphics[width=1.07\textwidth]{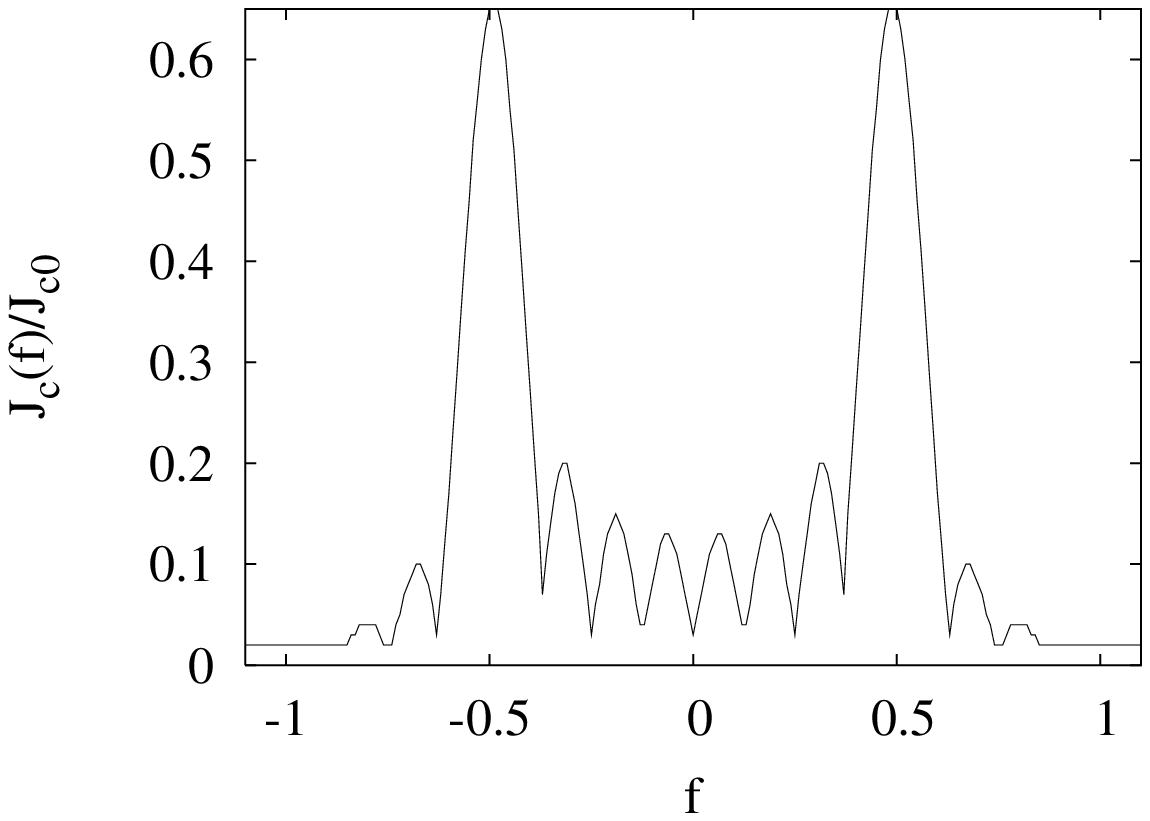}
\caption{\small Same as in Fig.\ \ref{eight_1mini}, but with five
mini-junctions per facet.  }\label{eight_5mini}
\end{center}
\end{figure}

\begin{figure}
\begin{center}
\includegraphics[width=1.07\textwidth]{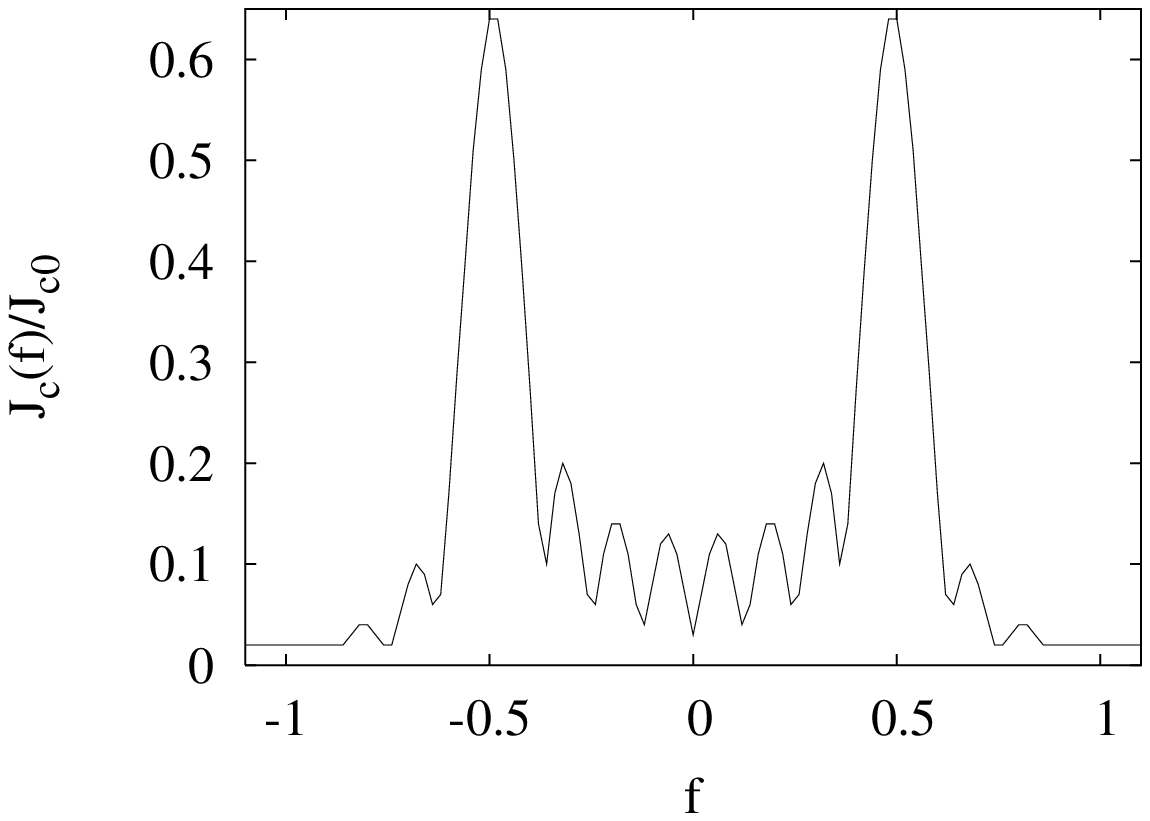}
\caption{\small Same as in Fig.\ \ref{eight_1mini}, but with ten
mini-junctions per facet.  }\label{eight_10mini}
\end{center}
\end{figure}

In Figs.\ \ref{eight_1mini}, \ref{eight_5mini}, and
\ref{eight_10mini}, we show the calculated $J_c(f)$ (obtained on the
increasing current branch) for a ladder consisting of eight
alternating $0$ and $\pi$ facets, open boundary conditions, $Q_J = 6$,
and $\lambda_J = 2.6 L_f$, but with each facet divided into one, five,
and ten mini-junctions, respectively.
%In Fig.\ \ref{eight_1mini} has many similarities to the experiment.
%For example, the experimental results show maxima in the critical
%current at $-0.5$ and $0.5\mu$T; at these maxima, the critical
%current (on the increasing current branch) is around $2\mu$A - very
%close to the measured {\em zero-field} critical current maximum of
%$2.2\mu$A.
In all three cases, our calculations show $J_c(f)$ to be symmetric
around $f = 0$, with maxima at $f = \pm 1/2$.  In the case $\Delta =
L_f$, $J_c(f = \pm 1/2)$ equals the zero-field critical current of a
pure $0$ junction, but for $\Delta = L_f/5$ and $\Delta = L_f/10$.
$J_c(\pm 1/2)$ is smaller than this value.  This behavior can be
understood as follows.  For $\Delta = L_f$, the eight-facet $0$-$\pi$
ladder becomes perfectly unfrustrated at $f = \pm 1/2$, and thus
should behave like the $f = 0$ ladder of $0$ junctions.  However, in
the other two cases, there is still some residual frustration at $f =
\pm 1/2$, and hence $J_c(\pm 1/2)$ is reduced.

$J_c(\pm 1/2)$ has approximately the same magnitude in Figs.\
\ref{eight_5mini} and \ref{eight_10mini}, suggesting that the $N_p =
5$ case is already close to the asymptotic limit ($\Delta \rightarrow
0$), at least for $f = \pm 1/2$.  In general, $J_c(f)$ has strong
peaks at $f = 1/2 + m$, where $m$ is an integer, but these peaks are
unequal except for $N_p = 1$.  From Figs.\ \ref{eight_5mini} and
\ref{eight_10mini}, we see that the heights of the first {\em three}
peaks in $I_c(f)$ do not change when we go from $N_p = 5$ to $N_p =
10$.  This is once again a consequence of the fact that $N_p = 10$ or
even $N_p = 5$ is close to the continuum limit for these values of $f$
and $N_f$.  The model calculations of Figs.\ \ref{eight_5mini} and
\ref{eight_10mini} are clearly more realistic than that with $N_p =
1$, because they allow $\phi$ to vary {\em within} a single facet.

%Another similarity is that experiment and calculation both give a
%critical current which is symmetric about zero applied field.  One
%difference is that at zero field, the $0$-$\pi$ ladder is calculated
%to have a relative minimum in the critical current, whereas in the
%experiment, there is a local maximum at $H_a = 0$. We speculate that
%these differences between calculation and experiment may arise from
%geometrical differences between the two systems(i. e. the absence of
%corners in our model), as mentioned earlier.

In Fig.\ \ref{eight_1mini}, but not Figs.\ \ref{eight_5mini} and
\ref{eight_10mini}, $J_c(f)$ is periodic with period unity.  This
periodicity is an artifact; it occurs because $N_f = 1$ in Fig.\
\ref{eight_1mini}, which causes the equations of motion to be periodic
with period unity.  By contrast, this periodicity is absent if $N_f >
1$.  If $L_f = N_p\Delta$, the equations of motion are periodic in $f$
with period $N_p$.  But since the physically relevant limit is $N_p
\rightarrow \infty$ or $\Delta \rightarrow 0$, this periodicity is
also an artifact of the discretization.

Another notable feature of Figs.\ \ref{eight_1mini}-\ref{eight_10mini}
is that each shows exactly six secondary maxima between the two
principal maxima in $J_c(f)$. These maxima occur at about the same
values of $f$ no matter how finely the facet is discretized. Moreover, they
do not start to overlap as the number of mini-junctions is increased.  The
maxima can be understood from a generalization of the Fraunhofer
pattern described above to the case of $N_f$ alternating $0$ and $\pi$
junctions. This generalization has been given in eq.\ (3) of Ref.\
\onlinecite{smilde}, and, using a slightly different derivation, in Ref.\
\onlinecite{mints}.  This generalized Fraunhofer pattern (not shown in the
Figure) agrees remarkably well with the numerically calculated
$J_c(f)$ our eight-facet junction.  We believe the slight deviation is
due to the fact that the phases within each facet, in our dynamical
model, are obtained directly by solving the equations of motion for
the long junction rather than by a superposition argument.

%In general, if a facet contains $n$ mini-junctions, then these
%maxima occur at fields $f = \pm n/2$, $\pm 3n/2$, $\pm 5n/2$, ... If
%$n = 1$, all these maxima are equal in height, because for this
%$I_c(f)$ is always periodic in $f$ with periodicity $1$.  However,
%for other values of $n$, the maxima have different heights, because
%the vortex configurations for different $n$ are not equivalent.

Finally, we comment on another feature of Figs.\ \ref{eight_5mini} and
\ref{eight_10mini}, but not Fig.\ \ref{eight_1mini}, namely the
existence of an approximate, but not exact, plateau in $J_c(f)$. In
our numerical studies for these two ladder systems, we find that the
smallest critical current density $J_c(f)$ is in the range of $1\% -
2\%$ of the critical current density $J_{c0}$ of a single facet.
Moreover, $J_c(f)$ remains in this range over a broad range of $f$.
This range of values of $1\% - 2\%$ remains the same regardless of how
small we make the incremental increases in the driving current.  We
believe that this approximate plateau can also be understood from the
superposition argument mentioned above.  More details of this
approximate plateau are discussed below.
%IVAN5: need more comments about the numerics here.  I plan to put these
%in.

%This $2\%$ effect is clearly visible in the experiments of
%\cite{darminto}, but for a system with more facets, namely, a zigzag
%junction with $80$, $5\mu$m facets.

\begin{figure}
\begin{center}
\includegraphics[width=0.75\textwidth]{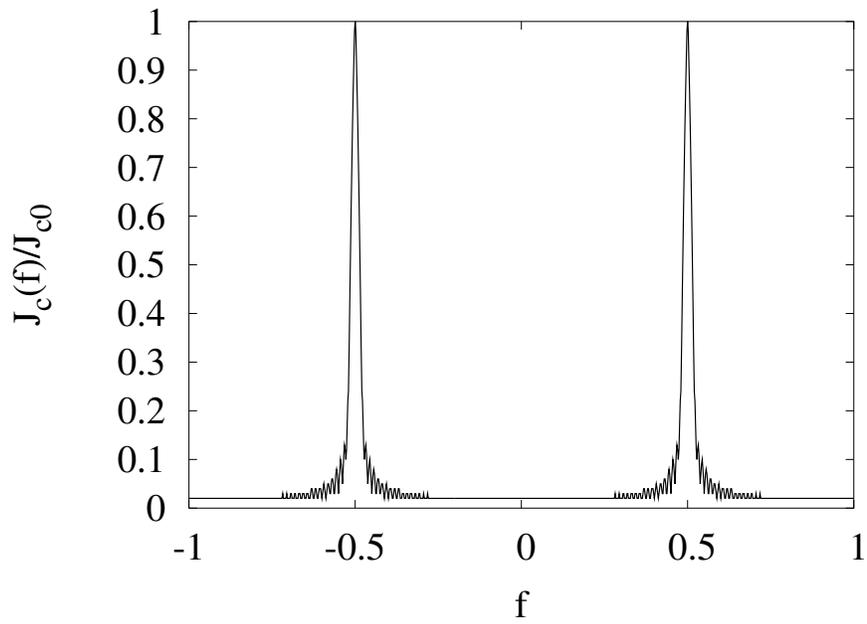}
\caption{\small Same as in Fig.\ \ref{eight_1mini}, but with 80
  facets, one mini-junction per facet, and $\lambda_J = 13
  L_f$.}\label{eighty_1mini}
\end{center}
\end{figure}

\begin{figure}
\begin{center}
\includegraphics[width=0.87\textwidth]{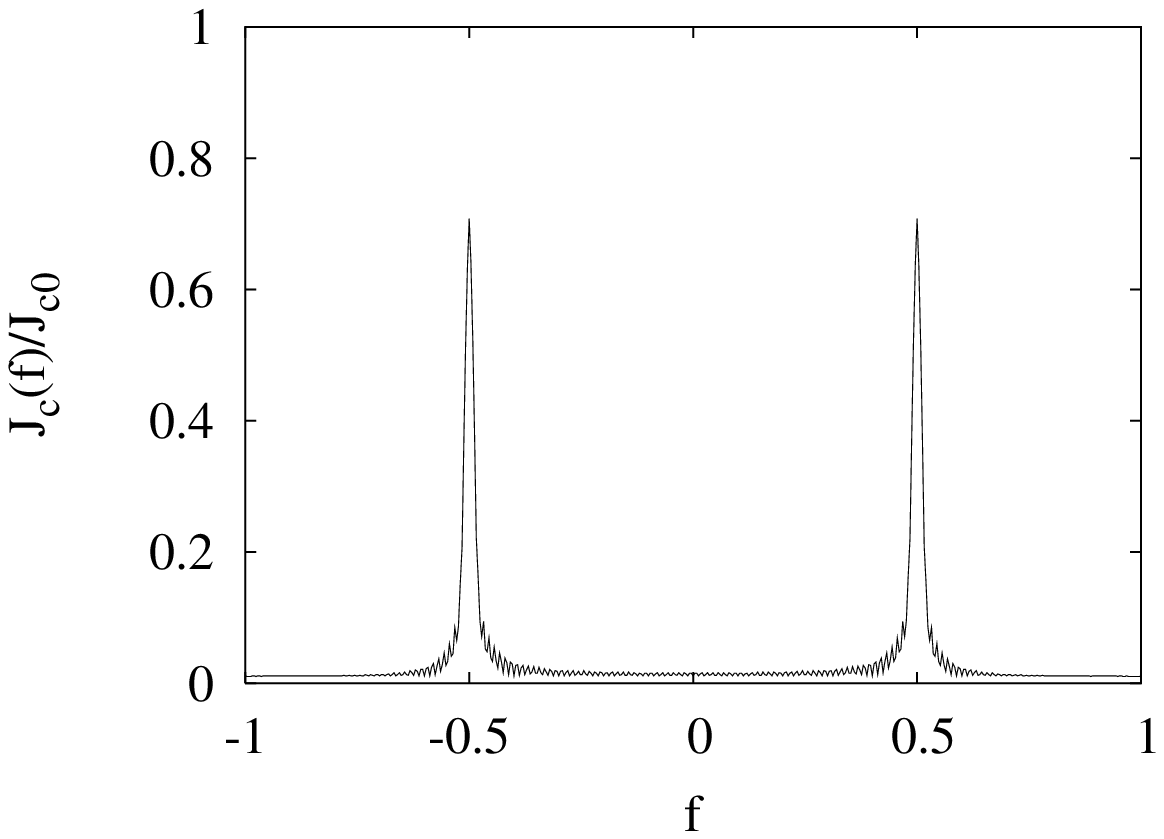}
\includegraphics[width=0.87\textwidth]{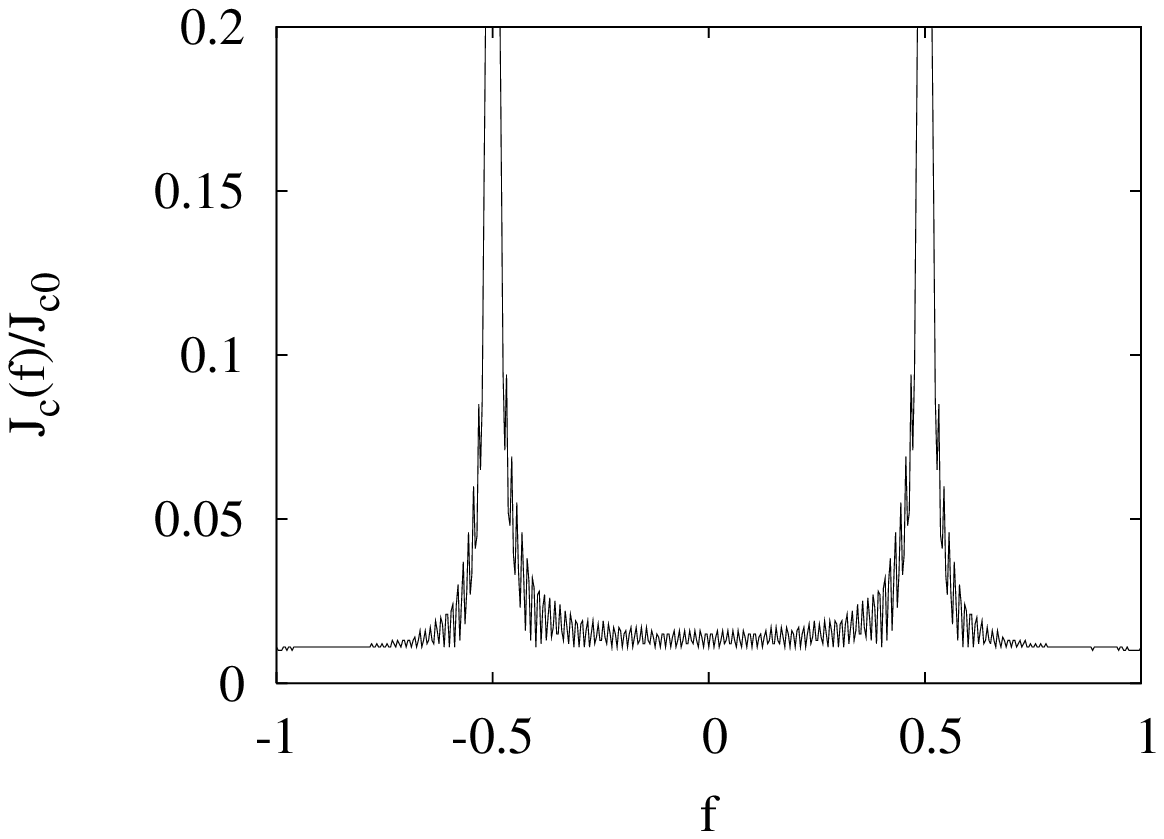}
\caption{\small  (a) (top): Same as in Fig.\ \ref{eighty_1mini}, but with two
mini-junctions per facet.  In contrast to Fig.\ \ref{eighty_1mini}, we have incremented the current density
in increments of $0.001J_{c0}$ to calculate the critical current.  With this fine scale, small oscillations in $J_c(f)$
are clearly apparent in the ``plateau'' region, which is not perfectly flat.   (b) (bottom):  same as (a), except that we have magnified
the vertical scale to make the oscillations more visible.}\label{eighty_2mini}
\end{center}
\end{figure}

\begin{figure}
\begin{center}
\includegraphics[width=0.87\textwidth]{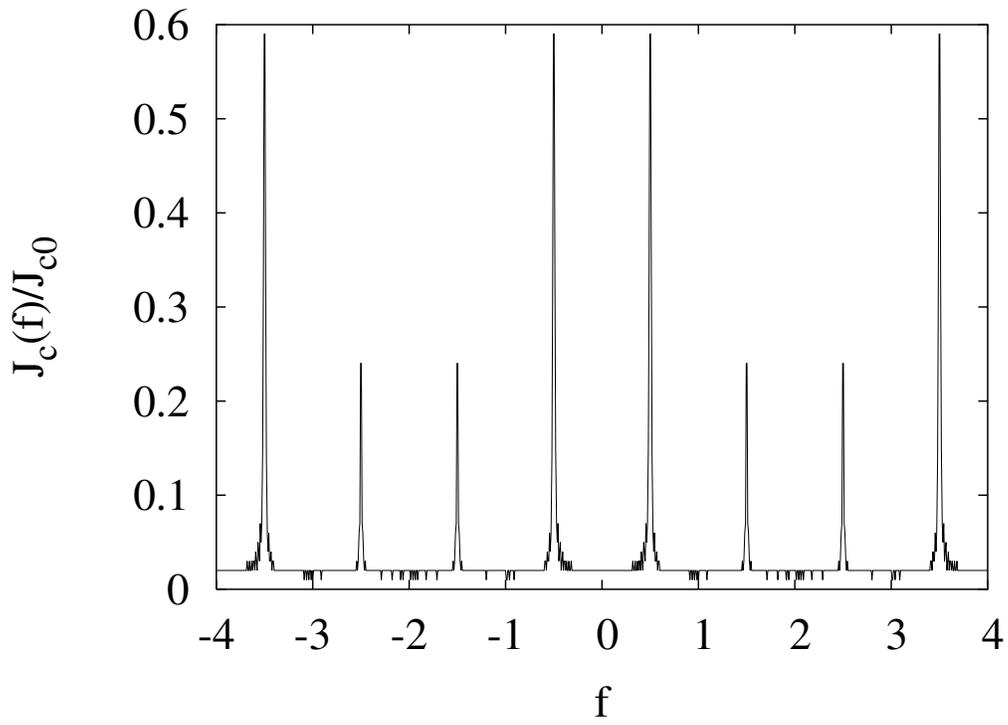}
\caption{\small Same as in Fig.\ \ref{eighty_1mini}, but with four
mini-junctions per facet.  As in Fig.\ \ref{eighty_1mini}, we increment the current density in intervals
of $0.01J_{c0}$.  At this resolution, most of the plateau region looks flat, but we expect that, if calculated
with increments of $0.001J_{c0}$ it would show oscillations as in Fig,\ \ref{eighty_2mini}.}\label{eighty_4mini}
\end{center}
\end{figure}

To further investigate the behavior of $J_c(f)$ in a different range
of the ratio $\lambda_J/L_f$, we have carried out additional
calculations for a ladder of alternating $0$ and $\pi$ facets with
$Q_J = 6$, $N_f = 80$, and a ratio $\lambda_J/L_f = 13$, the latter
two parameters being the same as in recent experiments\cite{darminto}.
We consider three cases: $N_p = 1$, $N_p = 2$, and $N_p = 4$.
%%This configuration is similar to the
%zigzag junction studied experimentally.
The results of these calculations, shown in Figs.\ \ref{eighty_1mini},
\ref{eighty_2mini}(a) and (b), and \ref{eighty_4mini}, are quite different from
those shown previously.  Now, for almost all values of $f$ between
$-0.5$ and $0.5$, $J_c(f) \sim 0.01 - 0.02J_{c0}$. Very close to $f =
\pm 1/2$, $J_c(f)$ increases slightly, and exactly at $f = \pm 1/2$,
$J_c(f)/J_{c0} = 1$ in Fig.\ \ref{eighty_1mini}. The behavior at $f =
\pm 1/2$ is consistent with expectations, since, for $N_p = 1$ the
ladder is unfrustrated at these values of $f$.  If we consider $N_p =
2$ and $N_p = 4$, but keep $\lambda_J/L_f = 13$, the positions of the
peaks remain unchanged and their heights change in a predictable
manner similar to that already found in
Figs.~\ref{eight_1mini}-\ref{eight_10mini}. 
% Figs.~\ref{eighty_2mini}
%and \ref{eighty_4mini} show such results for ladders with $N_f = 80$,
%$N_p = 2$ and $4$, with $\lambda_J = 13L_f$. 
In both Figures, as in
Fig.\ \ref{eighty_1mini}, there are, as expected, strong peaks in
$J_c(f)$ at $f = 1/2 +m$, with $m$ an integer. When $N_p \geq 2$, the
peak heights are smaller than unity, as already expected from the
results shown in Figs.~\ref{eight_5mini} and~\ref{eight_10mini}.  The
close resemblance between $N_p = 2$ and $N_p = 4$ suggests that the
continuum limit is already approached by $N_p = 4$.

We have carried out many other calculations (not shown here) for other
values of $\lambda_J/\Delta$.  In general, we find that this
approximate plateau occurs over the widest range of $f$ when
$\lambda_J/\Delta \gg 1$ and when $N_f \gg 1$.  Although the
experiments\cite{darminto} are not sensitive enough to measure
$J_c(f)$ in the range of $0.01$-$0.02 J_{c0}$, they do show that
$J_c(f) \sim 0$ over a broad range of $f$.

We have done various numerical checks to confirm that the plateau is
not a numerical artifact.  These checks are carried out primarily for the results of
Fig.\ 7.  To carry out the checks, we have typically,
for a given $f$, ramped up $J$ in units of 0.001$J_{c0}$, where
$J_{c0}$ is the critical current density of a single facet, rather than
0.01$J_{c0}$ as in the calculations shown in the other Figures. To
compute $J_c(f)$, after ramping up $J$, we iterated the coupled
Josephson equations for a dimensionless time $\omega_p t = 300$, then
averaged the voltage over the next interval of $200\omega_p t$.
$J_c(f)$ was taken as the value of $J$ for which the time-averaged
voltage, as determined in this way, first becomes non-zero. We then
incremented $f$ by $0.01$,
%IVAN6: What does ``non-zero'' mean here?  And how big is the increment xxx?
%Can you check the comments I made in the above paragraph?
%STROUD6:  Non-zero means here that it has a value that is greater than zero, that is
% to say numerically, it has a value that is greater than 1.0e-4.  As far
%as the value of the f increment at that is 0.01
reset $J$ to zero, and again incremented the current density in steps
of 0.001$J_{c0}$ to find the next $J_c(f)$.  In this way, we found
that the $J_c(f)$ in the plateau region varies between about
0.012$J_{c0}$ and 0.016$J_{c0}$.  

The results of these calculations, for a ladder of 80 facets with two mini-junctions per facet, are shown in Fig.\ \ref{eighty_2mini} (a) and (b).  
%IVAN10: I modified the sentence below.
Here, we see that $J_c(f)$, though small
over a wide range of $f$, is not actually perfectly flat, as mentioned above, but instead has small-amplitude oscillations as a function of $f$.   
%IVAN10: I have just now (August 22) added the following four sentences and modified the rest
%of this paragraph.
Note that, because of the amount of computer time needed to do these simulations in which
the current is incremented by only 0.001$J_{c0}$  
%IVAN10: do you remember if the increment is 0.001J_{c0} or 0.002J_{c0}?
rather than 0.01$J_{c0}$, we have done the small-increment calculations
only in Figs.\ 7(a) and 7(b), and not Figs.\ 6 and 8.   
As in all the other Figures showing $Jc(f)/J_{c0}$ versus $f$, $J_c(f)$ is taken to be
that value of $J$ above which the time-averaged voltage
$\langle V\rangle$ jumps up to its value on the resistive branch of
the hysteretic current-voltage characteristic, i. e. $\langle V\rangle
= J_{c0}/G$, where $G$ is the shunt conductance per unit length.
Clearly, $J_c(f)$ is indeed non-zero on the plateau, and remains nonzero over
a broad range of $f$ whenever $N_f$ is large enough.

\begin{figure}
\begin{center}
\includegraphics[width=0.90\textwidth]{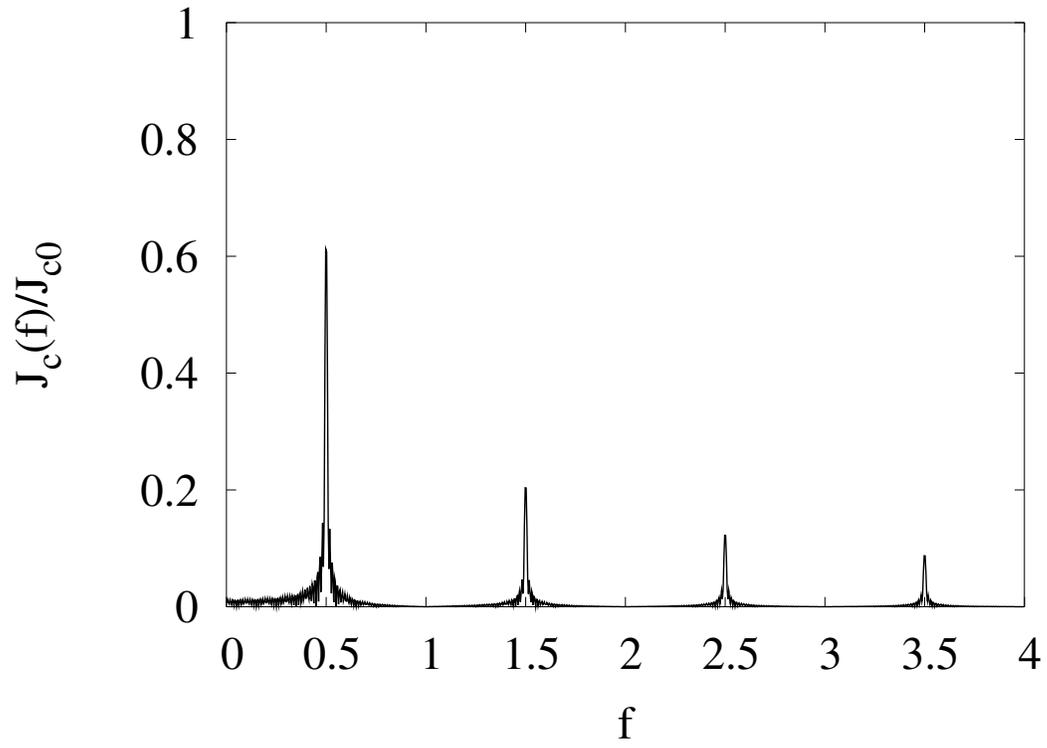}
\caption{\small $J_c(f)/J_{c0}$ as obtained from the superposition
approximation of Refs.\ \cite{smilde} and \cite{mints} for a long
$0$-$\pi$ junction having 80 facets}\label{smildemints}
\end{center}
\end{figure}

We have also calculated $J_c(f)$ using the simple superposition theory
of Refs.\ \cite{smilde} and \cite{mints}.  The results, shown in Fig.\
\ref{smildemints}, are remarkably similar to those in Fig.\
\ref{eighty_4mini}.  Thus, although our calculations are based on a
numerical solution of the coupled Josephson equations appropriate to
this geometry, the field-dependent critical current closely resembles
that obtained by a simple superposition theory.  In particular, the
heights of the first two sharp peaks in $J_c(f)$ as calculated
numerically, for $f > 0$, are similar to those obtained using the
generalized superposition approximation.

\begin{figure}
\begin{center}
\includegraphics[width=0.75\textwidth]{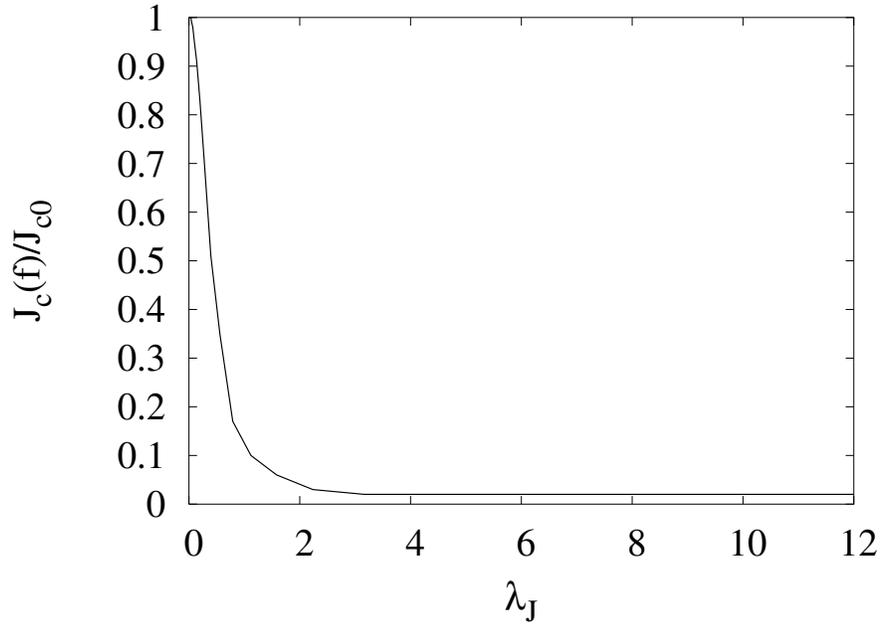}
\caption{\small Critical current per unit length $J_c(f)$, in units of
$J_{c0}$, for $f = 0$ and an $80$-facet $0$-$\pi$ junction modeled as
in the text for $f = 0$, assuming one mini-junction per facet, and
plotted as a function of $\lambda_J/L_f$.}\label{lambda_var}
\end{center}
\end{figure}

Finally, to show explicitly how $J_c(f)/J_{c0}$ depends on
$\lambda_J$, we have calculated the critical current at {\em zero}
applied field ($f = 0$) for an alternating $0$-$\pi$ ladder with 80
facets, each of length one junction ($N_p = 1$), as a function of
$\lambda_J$. The results are shown in Fig.\ \ref{lambda_var}.  The results of
this Figure are calculated incrementing $J/J_{c0}$ by 0.01 for each value of $f$.  
For very small $\lambda_J$ the ladder just behaves like many uncoupled
small junctions, and $J_c(0) = J_{c0}$.  As $\lambda_J$ increases,
$J_{c0}$ decreases, until $\lambda_J \sim \sqrt{10}$.  For $\lambda_J
> \sqrt{10}$, $J_c(0) \sim 0.02 J_{c0}$.  Once again, this value is
not exact, but fluctuates slightly with $\lambda_J$.

\section{Generalized Superposition Limit}

%THIS SECTION MODIFIED STARTING 09 APRIL 2007

Since our results closely resemble the simple superposition theory (or
``generalized Fraunhofer limit''), we have examined our equations of
motion [eqs.\ (\ref{eq:phiddot}) - (\ref{eq:phiddotn})] in an effort
to understand when that model gives the superposition limit.  
First, we sum these equations to obtain an expression for
$(1/N)\sum_{i=1}^N\ddot{\phi_i}$.  With the definition $\phi =
(1/N)\sum_{i=1}^N\phi_i$, and the use of several trigonometric
identities, we finally obtain
\begin{equation}
\ddot{\phi} = - A\sin(\phi + \theta) + J/J_{c0} - \dot{\phi}/Q_J, 
%+[\lambda_J^2/(N_f\Delta)^2](\phi_N - \phi_1)/N, 
%This last term is actually not present - it was included because of an algebraic error
%in our last draft.
\label{eq:super1}
\end{equation}
where
\begin{eqnarray}
A & = & \sqrt{C^2 + S^2}, \\
\theta & = & \tan^{-1}(S/C), \\
C & = &\frac{1}{N}\sum_i\cos(2\pi i f/N_p + n_i\pi + \phi_i - \phi), \\
S & = & \frac{1}{N}\sum_i\sin(2\pi i f/N_p + n_i\pi + \phi_i - \phi).
\end{eqnarray}

The ``superposition limit'' is appropriate when the phase $\phi_i$ is independent of $i$.  We expect this to be true when $\lambda_J \gg L_f$, in which case $\phi_i$ should vary little across the
zigzag junction.  In this limit, $\phi_i - \phi = 0$, both $C$ and $S$ are independent of $\phi$, and eq.\ (\ref{eq:super1}) just becomes the equation of motion for a single RCSJ
with a field-dependent critical current and quality factor.  Both of these can be obtained
with the change of variables $\psi = \phi + \theta$, in terms of which
eq.\ (\ref{eq:super1}) becomes
\begin{equation}
\ddot{\psi} = -A\sin(\psi) + J/J_{c0} -\dot{\psi}/Q_J.
\end{equation}
With the further change of variables $\tau^\prime = \tau/\sqrt{A}$, this
last equation becomes
\begin{equation}
\ddot{\psi} = -\sin\psi + J/(AJ_{c0}) - \dot{\psi}/(\sqrt{A}Q_J).
\label{eq:super2}
\end{equation}
Eq.\ (\ref{eq:super2}) describes a single Josephson junction with a critical
current density 
\begin{equation}
J_c^s(f) = AJ_{c0}
\end{equation}
and quality factor 
\begin{equation}
Q_J^s(f) = 
\sqrt{A}Q_J = Q_J\left(\frac{J_c(f)}{J_{c0}}\right)^{1/2},
\label{eq:qjs}
\end{equation}
where the superscript denotes the superposition approximation.

To obtain the value of $A$ (in the physically relevant continuum limit
$\Delta \rightarrow 0$), we rewrite $C$ as 
$C = \mathrm{Lim}_{N\rightarrow
\infty}(N\Delta)^{-1}\sum_{i=1}^N[\Delta\cos(2\pi i f/N_p + n_i\pi]$
or, converting the sum to an integral, $C = 1/(N_fL_f)\int_0^{N_f}L_f
dx\cos[2\pi f x/L_f+n(x)\pi]$.  The integral can be done analytically.
$S$ can be obtained in analogous fashion.  The resulting expressions
for $C^2$ and $S^2$, and hence $A = \sqrt{C^2 + S^2}$ and $J_c^s(f) =
AJ_{c0}$, can be shown to reduce to that given in eq. (3) of Ref.\
\onlinecite{smilde}, namely (using our variables as defined earlier)
\begin{equation}
C^2 = \frac{1}{2\pi f}\left\{\sum_{n = 1}^{N_f}\sin\left[n(\pi - 2\pi f)\right][1 - \cos(2\pi f)] 
-  \sum_{n=1}^{N_f}\cos\left[n(\pi - 2\pi f)\right]\sin(2\pi f)\right\}^2
\end{equation}
and
\begin{equation}
S^2 = \frac{1}{2\pi f}\left\{\sum_{n=1}^{N_f}\cos\left[n(\pi - 2\pi f)\right][1 - cos(2\pi f)]
+ \sum_{n=1}^{N_f}\sin\left[n(\pi - 2\pi f)\right]\sin(2\pi f)\right\}^2
\end{equation}

Thus, our approach does reduce, as it should, to simple superposition in the limit
when the phases $\phi_i$ are independent of $i$.  We expect this limit to be appropriate
when $\lambda_J \gg L_f$.    We also obtain an additional piece
of information in this limit, namely, the effective value of the
junction quality factor $Q_J^s(f)$.  Our result for this quality factor is not particularly
surprising.  The reason why the quality factor becomes smaller at certain fields is not that
the local damping changes with field.  Rather, it is because the
critical current is reduced at certain magnetic fields by cancellation between different
parts of the junction, leading to an increase in the {\it relative} dissipation in the
junction at these fields.  If $\lambda_J/L_f$ is finite, the superposition approximation is
not exact, because $\phi_i$ will be dependent on $i$.   In this case, we should expect deviations from its predictions for $J_c(f)$.  These are, indeed, observed in our calculations 
(and shown explicitly in Fig.\ 10).

\section{Discussion}

We have presented a simple model for field-dependent critical current
density $J_c(H_a)$ of a long $0$-$\pi$ junction.  An important example
of such a system is a zigzag junction connecting an $s$-wave
superconductor and one with a $d_{x^2-y^2}$ order parameter.  We
numerically solve for the IV characteristics of this junction by
discretizing the equation of motion in space and time.  The model
gives, not only $J_c(H_a)$, but also (in principle) the full IV
characteristics and its hysteretic behavior when the shunt resistivity
is large.

In the limit $\lambda_J \rightarrow \infty$, $J_c(H_a)$ for our model
reduces to that predicted by a generalized superposition
model\cite{darminto,mints}.  However, our model is more general,
because it applies for finite $\lambda_J/L_f$ and hence accounts for
the variation of the phase along the junction.  Furthermore, even in
the limit $\lambda_J \rightarrow \infty$, the model gives not only the
analytical form for $J_c(H_a)$, as previously obtained in Refs.\
\cite{smilde} and \cite{mints}, but also a field-dependent effective
quality factor $Q_J(H_a)$.  The field-dependence of $Q_J(H_a)$ is due to cancellation
effects within the long $0$-$\pi$ junction as a function of $H_a$, rather than any field-dependence
of the assumed local damping within the junction. 

Our numerical results qualitatively resemble the experiments on
$d$-$s$ zigzag junctions\cite{smilde,darminto}.  As in the experiment,
we obtain a strong maximum in $J_c(H_a)$ at a non-zero magnetic field,
which is comparable in magnitude to its unfrustrated value, $J_{c0}$.
Furthermore, our results, like the experimental ones, depend strongly
on the ratio $\lambda_J/L_f$.  Finally, for large $\lambda_J/L_f$ and
large numbers of facets $N_f$, we we find that $J_c(H_a)$ is very
small over a broad range of magnetic field and in the range of
$1$-$2$\% of $J_{c0}$.  The experiments are not sensitive enough to
measure such a small $J_c(H_a)$ but they also show broad ranges of
$H_a$ where $J_c(H_a)$ is very small.  The origin of this large region
of very small $J_c(H_a)$ is undoubtedly the cancellation resulting from
superimposing the currents from different parts of the long $0$-$\pi$
junction, as implied by the superposition model.  The experimental
results appear to have some fine structure not present in our model.
We do not know the origin of this fine structure, but speculate that
it may be due to the bends in the zigzag junction, which are not
included in the model.

Finally, we also comment briefly on the applicability of the RCSJ model to a junction
connecting an s-wave superconductor to one with d$_{x^2-y^2}$ symmetry.  One possible
concern is that, in the d-wave superconductor, since the energy gap vanishes at certain
points in k-space, it is relatively easy for a current to excite quasiparticles and, therefore,
possibly heat the superconductor.  Therefore, if a fluxon moves through a long $0$-$\pi$ junction, 
it might easily cause localized heating in the junction.    If this heating occurs,
it would tend to mask the RCSJ behavior of the long $0$-$\pi$ junction.  One might also
ask how accurately the RCSJ model would describe the long junction, even without heating.
Clearly, the RCSJ model is an idealized approximation of the real junction behavior.  But
the model seems to be a reasonable starting point for possible, more refined approaches.
Moreover, the RCSJ behavior can potentially be tested, in the limit
when the Josephson penetration depth $\lambda_J$ is large compared to a facet length.
Namely, in this limit, the RCSJ model implies that the effective 
junction quality factor $Q_J(H_a)$ is proportional to the square root of the field-dependent
critical current density $J_c(H_a)$.  If experiments show this dependence, this would represent
evidence that the RCSJ model is indeed applicable to a long $0$-$\pi$ junction.

The present work suggests several questions which could be further
studied in experiments.  For example, it would be of interest to study
$0$-$\pi$ zigzag junction when $\lambda_J/L_f$ is smaller than unity.
In this range, we would expect substantial departures from the
generalized superposition model, as suggested by Fig.\
\ref{lambda_var} for $f = 0$. In addition, it would be desirable to
test the predicted dependence of $Q_J(f)$ on $f$ experimentally in the
superposition limit.  Since $J_c(f)$ may be small over a broad range
of $f$, eq.\ (\ref{eq:qjs}) suggests that a long $0$-$\pi$ junction
would behave as if overdamped at these values of $f$ even if it is
underdamped and hysteretic when $J_c(f)$ is large.  It would be of interest to
test this hypothesis experimentally.

\section{Acknowledgments.}  This work has been supported through NSF
Grant DMR04-13395.  Some of the calculations were carried out using
the facilities of the Ohio Supercomputer Center, with the help of a
grant of time.

\newpage

%IVAN6: Again, need to change vertical axis to J_c(f)/J_{c0}.  Also, if possible,
%it would be best to repeat this for several more periods of $f$, so that it can
%be compared to Figs. 4 and 5, where the strong peaks are not of equal height.  
%Also, all three of Figs. 3, 4, and 5 should be on the same scale, if possible,
%for ease of comparison.  The second referee suggested this.
%Maybe you could go out, say, far enough to include four peaks in each case?
%STROUD6: I only ran all three runs from -0.8 to 0.8 f, I have redone the scales
%on the figures, but I don't have data past f = 0.8.  I can rerun them if you want
%me to.  

%IVAN6: In Figs. 6, 7, and 8, for ease of comparison, can you include three
%strong peaks in the region f > 0?  Also, once again change the vertical
%axes to J_c(f)/J_{c0} in all three Figures.  Finally, can you redefine the
%horizontal axis so that $f$ means the flux per facet in all three of these
%Figures (as in the earlier set of three).
%STROUD6: I changed the ranges on the vertical axes and changed the notation
%on the vertical axes on figures 7 and 8, however I again don't have data
%to include more peaks, but I can run more if you would like me to.  

%IVAN6: I thought of adding your finer-scale Figure as part (b) here.  Another alternative might
%be to replace the entire Figure 7 by your new one with a finer scale, but this would look a little
%strange when Figs. 6 and 8 are computed using the old increments.  What do you think?
%STROUD6:  I agree it would look strange to just replace it.  We don't mention anywhere in the text
%running finer-scale calculations.  So, should we even put it in here at all?  I'm just not sure
%we should add it.  I think that you made it clear in the text that it is not a constant 2%.  

%IVAN6: I am sending the data for this calculation as a Table.  Would you be willing to plot it?
%I think you have better plotting software than I do. 
%STROUD6: This has been done and should have been attached with the email that I sent this in.

\end{document}